\documentclass[]{aa}
\usepackage{graphicx}
\begin{document}
%\thesaurus{4(11.13.2, 11.19.2, 13.18.1, 02.16.2)}
\title{The large scale magnetic field structure of the spiral galaxy NGC\,5775}
\author{ Marian Soida \inst{1}
\and Marita Krause \inst{2}
\and Ralf-J\"urgen Dettmar \inst{3}
\and Marek Urbanik\inst{1}}
 
\institute{Astronomical Observatory, Jagiellonian University, ul. Orla 171,
30-244 Krak\'ow, Poland\and Max-Planck-Institut f\"ur Radioastronomie, Auf dem H\"ugel 71, 53121 Bonn, Germany 
\and Astronomisches Institut der Ruhr-Universit\"at Bochum, Universit\"atsstr. 150, 44780 Bochum, Germany }
 
\offprints{M. Soida}
\mail{soida@oa.uj.edu.pl}
 
\date{Received 07/08/2008; accepted 25/05/2011}
 
\titlerunning{The large scale magnetic field of NGC\,5775}
\authorrunning{M. Soida et al. }
 
\abstract
%context 
{The origin of large-scale magnetic fields  in spiral galaxies is still a
  theoretical riddle and better observational constraints are required to make
  further progress.
}
%aims
{
In order to better determine the large-scale 3D-structure of magnetic fields 
in spiral galaxies
we present a Faraday rotation analysis of the edge-on spiral galaxy NGC\,5775.
}
%methods
{
Deep radio-continuum observations in total power and linear polarization were
performed at 8.46\,GHz with the VLA and the 100-m Effelsberg telescope. They
were analyzed together with archival 4.86 and 1.49\,GHz VLA-data. We thus
can derive rotation measures from a comparison of three frequencies and
determine the intrinsic magnetic field structure.
}
%results
{
A very extended halo is detected in NGC\,5775, with magnetic field lines
forming an X-shaped structure. Close to the galactic disk the magnetic
field is plane-parallel. The scaleheights of the radio emission
esimated for NGC\,5775 are comaprable with other galaxies.
The rotation measure distribution
varies smoothly on both sides along the major axis from positive
to negative values. 
}
%conclusions
{
From the derived distribution of rotation measures and the plane-parallel
intrinsic magnetic field orientation along the galactic midplane we conclude
that NGC\,5775 has an \emph{even axisymmetric} large-scale magnetic
field configuration in the disk as generated by an $\alpha\Omega$-dynamo
which is accompanied by a quadrupolar poloidal field. The magnetic field
lines of the plane-parallel component are pointing \emph{outwards}.
The observed X-shaped halo magnetic field, however, cannot be explained
by the action of the disk's mean-field dynamo alone. It is probably due to the influence of the galactic wind together with the dynamo action.
}
 
 
\maketitle
 
\section{Introduction}
 
The large scale structure of magnetic fields in spiral galaxies is a strong
constraint for theories of the origin and evolution of cosmic magnetic fields.
For external galaxies the best tracer of the magnetic field is the
polarized synchrotron emission from cosmic ray electrons.
For the understanding of the large scale structure of 
galactic magnetic fields in disk galaxies the halo field is of significant
importance. In particular in view of the dynamo models which use the disk-halo
interaction of the interstellar medium (ISM) to explain the growth of magnetic
fields. 
 
Due to the
origin and propagation properties of cosmic ray electrons the halo magnetic
field is best traced in galaxies with strong star formation in the disk. 
In these galaxies the star formation drives the disk-halo 
interaction which also
allows cosmic ray electrons to escape into the halo (Dahlem et al. \cite{da06}).
The propagation of cosmic ray electrons into the halo may in turn contribute
to the growth of the magnetic field on large scales via e.g. fast dynamo
processes (Hanasz \& Lesch~\cite{HL03}).
 
One such galaxy showing all signatures of the disk-halo interaction is the
edge-on spiral NGC\,5775. X-ray emission from an extended halo of hot gas 
(T\"ullmann et al \cite{tul06}) associated with a thick disk of warm ionized 
H$^+$ gas (e.g. Rand~\cite{rand00}, Rossa \& Dettmar \cite{RD00}) and a halo
of synchrotron radio-continuum (Duric et al. \cite{dur98}) can be traced
up to $\ge$8\,kpc out of the plane. 
 
The disk-halo interface in this object has been extensively studied (e.g. Lee
et al. \cite{lee02}) and in particular correlations between the
radio, H$\alpha$, and \ion{H}{i} spurs (Collins et al.~\cite{col00}) 
raised speculations
about the role of magnetic fields in the propagating cosmic rays and the
transport of ionized gas from the disk into the halo. 
The inclusion of magnetic reconnection (Birk et al.~\cite{birk98}) could also
greatly relax problems with heating and ionization of the extraplanar gas
(e.g. Collins \& Rand \cite{CR01b}).
 
In a previous study  (T\"ullmann et
al.~\cite{tul00}) archival VLA data at 4.86 and
1.49\,GHz were used to construct polarization maps showing the magnetic field
structure in NGC\,5775. The radio halo of NGC\,5775 is found to be
substantially  polarized (polarization degree at 4.86\,GHz reaching 20\%
-- 30\%) with observed polarization B-vectors significantly inclined to
the plane (T\"ullmann et al.~\cite{tul00}). The polarized intensity was
found to make an X-shaped pattern extending along the direction of the
observed B-vectors.
 
\begin{table}[htbp]
\caption{Basic properties of NGC\,5775 (from LEDA database)}
\label{obs}
\begin{center}
\begin{tabular}[]{lc}
\hline
names& NGC\,5775\\
&UGC\,9579\\
R.A.$_{2000}$& $\mathrm{14^h53^m57.\!\!^s 6}$\\
Decl.$_{2000}$& $03^\circ 32^\prime 40.\!\!^{\prime\prime}0$\\
type& Sbc\\
incl.& 86$\degr^b$\\
pos.angl.&145$\degr$\\
distance& 26.7 Mpc$^a$\\
 & $1\arcmin \simeq 7.8$\,kpc \\
\hline
$^a$\,Dahlem et al.~\cite{dahl95}\\
$^b$\,Irwin~\cite{irwin94}
\end{tabular}
\end{center}
\end{table}
 
The interpretation of these observations has severe limitations: weak
polarized signal at 1.49\,GHz excludes a reliable
determination of Faraday rotation. In this work we present the 
reanalysis of the data in the context of new VLA observations at
8.46\,GHz, which allow us to account for the Faraday rotation.
 
\section{Observations and data reduction}
 
The radio observations at 4.85\,GHz were already described in  T\"ullmann
et al. (\cite{tul00}). New radio observations were made using the Very
Large Array interferometer (VLA) of the National Radio Astronomical
Observatory (NRAO\footnote{NRAO is a facility of National Science
Foundation operated under cooperative agreement by Associated
Universities Inc.}). The data at 8.46\,GHz were collected using the most
compact (D-array) configuration for the best sensitivity to extended
emission. Deep continuum observations (about 13 hours on source) with
full Stokes parameters were recorded.  Data at other
frequencies were taken from the NRAO VLA archive from  various
observation projects. A total of 10 hours on-source time at 4.86\,GHz (in
D-array), and 14 and 2.5 hours at 1.49\,GHz (in D, and C-array
respectively) were found useful for a polarimetry study.
 
All interferometric data were reduced using the AIPS data processing
package. The nearby point source 1442+101 was used for phase and 
instrumental polarization calibration. The source 3C\,286 was used for 
calibrating flux density scale and polarization position angle.
 
%====================================
\begin{table}[htbp]
\caption{R.M.S. noise level ($\sigma$) obtained in final total power and
polarized intensity maps}
\label{rms}
\begin{center}
\begin{tabular}[]{rrrrrrrl}
\hline
freq. & $\sigma$(total&\multicolumn{2}{c}{$\sigma$(Stokes}&$\sigma$(pol.&beam \\
&power)&\multicolumn{2}{c}{ Q and U)} & int.)&size\\
~[GHz]&[${\mu\mathrm{Jy}\over\mathrm{b.a.}}$]&\multicolumn{2}{c}{[${\mu\mathrm{Jy}\over\mathrm{b.a.}}$]}&[${\mu\mathrm{Jy}\over\mathrm{b.a.}}$]&[$\arcsec$]\\
\hline
8.46 & 13  &  7 & 8 & 9  & 16   & D-array\\
8.35 & 400 & 70 &72& 75 & 84   & Effelsberg\\
4.86 & 10  &  6 & 7 & 7  & 16   & D-array\\
1.49 & 60  & 23 &23 & 25 & 16   & C\&D-array\\
\hline
\end{tabular}
\end{center}
\end{table}
%====================================
 
%=======================================
\begin{figure}[htbp]
\resizebox{\hsize}{!}{\includegraphics[bb=40 155 560 730]{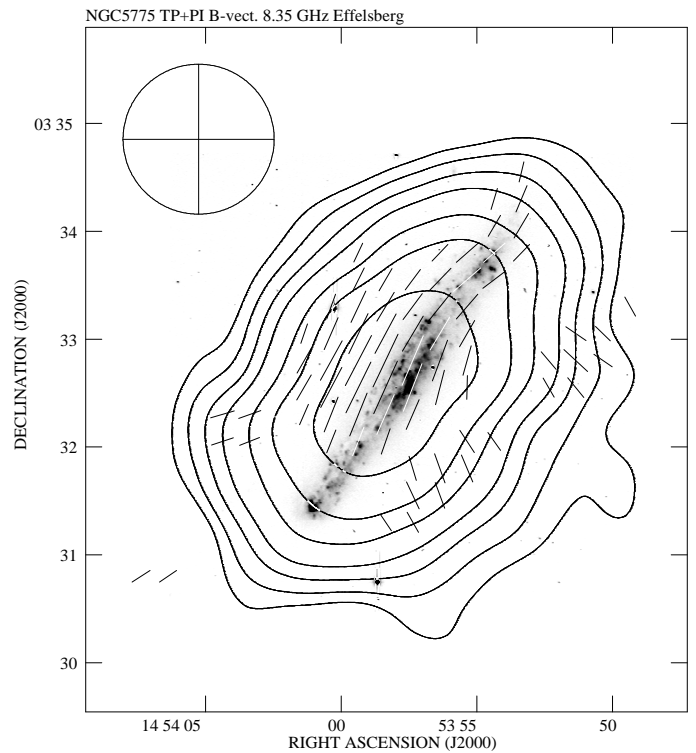}}
\caption[]{
Total intensity map at 8.35\,GHz (contours)with polarization observed
with the 100-m Effelsberg telescope, superimposed on the H$\alpha$
image (from T\"ullmann et al.~\cite{tul00}).
The contour levels are (3, 5, 8, 12, 20, 30, 50)$\times$
400\,$\mu$Jy/b.a. (r.m.s.). The vectors show the apparent B-orientation,
their lengths are proportional to the polarized intensity (1\arcmin{}
corresponds to 1.2\,mJy/b.a.). The angular resolution is 84\arcsec{} HPBW.
\label{eff}
}
\end{figure}
%===========================================
 
%=======================================
\begin{figure}[htbp]
\resizebox{\hsize}{!}{\includegraphics[bb=40 200 560 690]{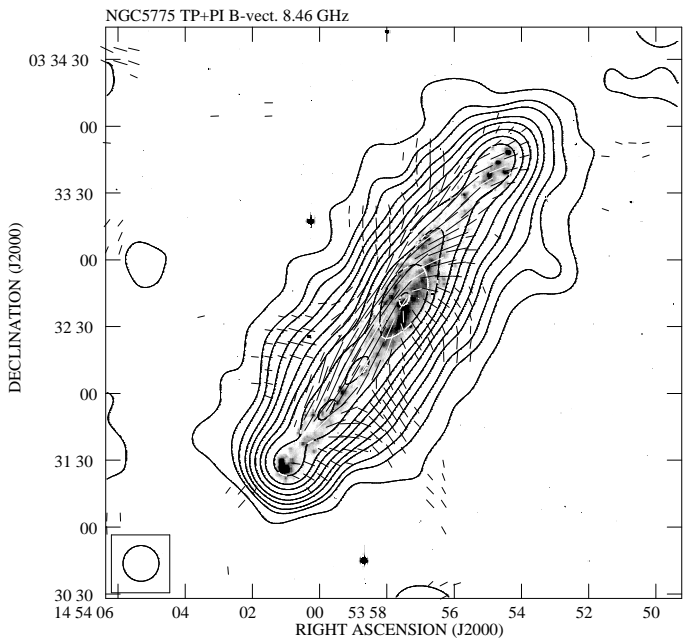}}
\caption[]{
Total intensity contour map with apparent polarization
B-vectors at 8.46\,GHz (combined VLA + Effelsberg)
superimposed on the H$\alpha$ image (from T\"ullmann et
al.~\cite{tul00}). Contours are plotted at levels (3, 5, 8, 12, 20, 30,
50, 80, 120, 200, 300 and 500)$\times$ 13$\mu$Jy/b.a.
(r.m.s.  level this frequency). Vectors are proportional to
the polarized intensity and indicate apparent B-orientation 
(10\arcsec{} corresponds to 100\,$\mu$Jy/b.a.).
The resolution is 16\arcsec{} HPBW.
\label{tp3}
}
\end{figure}
%===========================================
 
Complementary single-dish mapping of the
galaxy was made with the 100-m Effelsberg\footnote{
Effelsberg 100-m telescope is operated by
Max-Planck-Institut f\"{u}r Radioastronomie in Bonn (MPIfR)
on behalf of the Max-Planck-Geselschaft.}
radio telescope.
The Effelsberg $\lambda3.6$\,cm (8.35\,GHz) observations were made with
a single-beam receiver in the secondary focus of the 100-m telescope.
The receiver has two canals (RHC, LHC) with total-power
amplifiers and an IF polarimeter. The bandwidth was 1.1\,GHz, the system
noise temperature  about 25\,K, and the resolution 84\arcsec{} HPBW.
 
We obtained 33 coverages in total of NGC\,5775 between February and July 2003.
Each coverage has a map size of $18\arcmin \times 18\arcmin$ and was
scanned alternatively along R.A. and Dec. directions. The scanning
velocity was $30\arcsec/{\rm s}$ and the grid size is $30\arcsec$.
 
For the pointing and focusing we observed regularly the
source 3C\,286. The flux calibration was also done with 3C\,286 according
to the flux values of Baars et al.~(\cite{baa77}). All coverages were
combined (Emerson \& Gr\"ave~\cite{EG88}) separately for the Stokes
parameters I, Q, and U.  The rms~noise levels of the final maps are given
in Tab.~\ref{rms}.
 
%=======================================
\begin{figure*}[htbp]
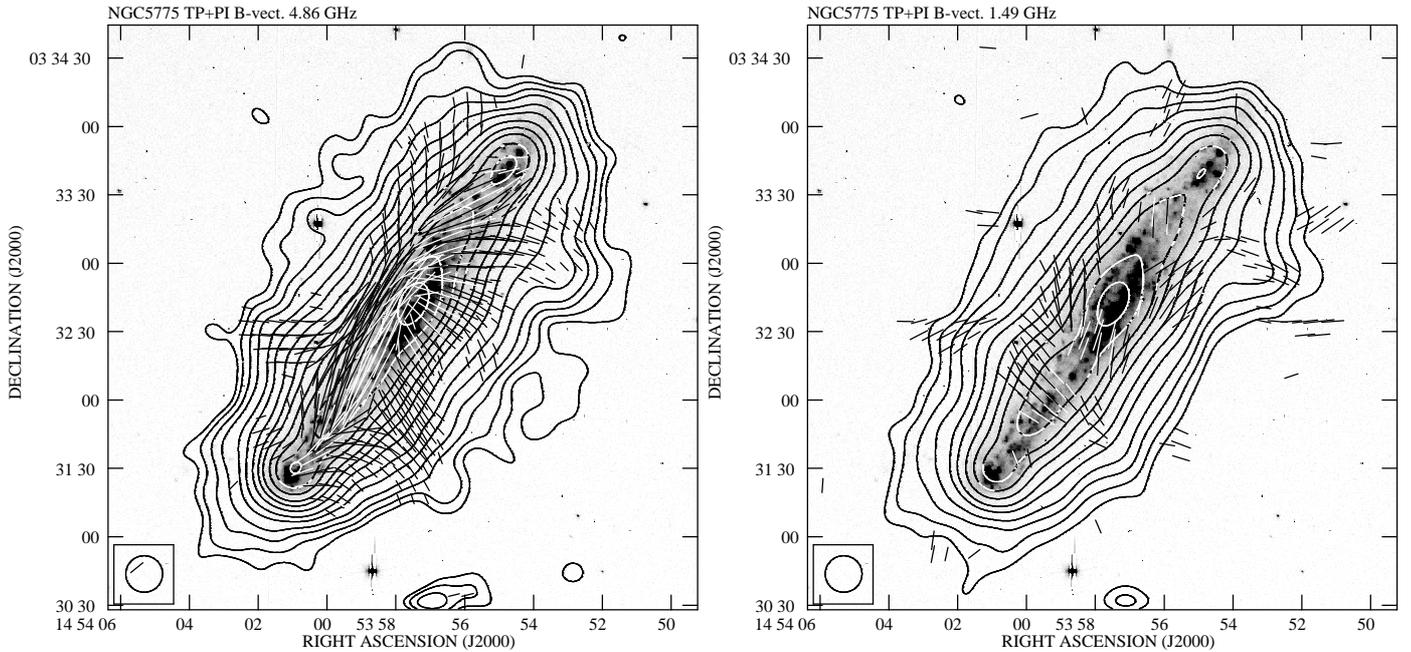

\resizebox{.5\hsize}{!}{\includegraphics[bb=40 200 560 690]{10763f03.ps}}%
\resizebox{.5\hsize}{!}{\includegraphics[bb=40 200 560 690]{10763f04.ps}}
\caption[]{
Total intensity contour map with apparent polarization
B-vectors at 4.86\,GHz (left) and 1.49\,GHz (right).
Both maps are superimposed on the H$\alpha$ image (from T\"ullmann et
al.~\cite{tul00}). Contours are plotted at levels (3, 5, 8, 12, 20, 30,
50, 80, 120, 200, 300 and 500)$\times$ 10 and 60\,$\mu$Jy/b.a.,
respectively (r.m.s.
levels at corresponding frequencies). Vectors are proportional to
the polarized intensity (10\arcsec{} corresponds to 50\,$\mu$Jy/b.a.)
and indicate apparent B-orientation.
The resolution is 16\arcsec{} HPBW in both maps.
\label{tp6_20}
}
\end{figure*}
%===========================================
 
The single-dish data at 8.35\,GHz were merged with the interferometric ones 
providing zero-spacing information, that is missing in the interferometric
data alone. The merging was performed in the Fourier plane (AIPS task IMERG)
with the Effelsberg map scaled in intensity to the VLA observing frequency,
assuming a constant spectral index of $\alpha=$0.85 ($S_\nu\propto\nu^{-\alpha}$).
About 3\% of the total intensity flux was recovered with this 
procedure. Complimentary single-dish observations were made also at 4.85\,GHz
using the 100-m Effelsberg telescope. Both single-dish and interferometric data
had the same integrated total flux and so no merging was performed at
this frequency.
 
The maps in all Stokes parameters (I, Q, and U) were convolved to
a common beam with HPBW of 16\arcsec{} and combined to obtain final maps of 
total and polarized intensities and polarization magnetic angles (E-vectors
rotated by 90\degr) called hereafter as ``apparent magnetic field
orientations'' (in short: apparent B-orientation).
 
\section{Results}
 
\subsection{Single-dish data}
 
The total intensity map obtained with the 100-m Effelsberg telescope
alone at 8.35\,GHz is presented in Fig.\ref{eff} along with polarization
B-vectors that give the apparent magnetic field orientation (i.e. without
correction for Faraday rotation, which is small at such a high frequency)
projected to the sky plane.
The intensity distribution shows a small asymmetry with respect to the
disk plane of the galaxy, being more extended to the south-western
side of the disk. Two extensions can be noticed on the 
southern and eastern side of the
galaxy (on both sides of the disk plane) both accompanied by polarization
magnetic vectors at an angle of about 45\degr{} to the plane. The apparent
magnetic field orientations close to the disk are plane parallel along
the whole radial extent of the disk. The 84\arcsec{} beam 
``smoothes''
the disk-parallel orientation of the magnetic vectors to the north-eastern
side of the disk plane.  Further away from the disk, the vectors
are nearly perpendicular to the disk in the western side of the disk plane.
The gap in the distribution of the polarization vectors
on the western part, between the two predominant orientations of the
vectors, is clearly caused by beam depolarization: the magnetic field
changes its direction by about 90\degr{} within the large single-dish beam.
 
\subsection{High-resolution total intensity data}
\label{sec:tp}
 
The total intensity distributions of NGC\,5775 at three radio wavelengths 
(8.46\,GHz,
4.86\,GHz and 1.49\,GHz) and the corresponding apparent magnetic vectors
(with lengths proportional to polarized intensity)  are presented in
Figs.~\ref{tp3}~\&~\ref{tp6_20} overlaid on the
H$\alpha$ distribution (from T\"ullmann et al.~\cite{tul00}). 
The distribution is quite similar at
all three frequencies, showing smooth maxima of emission clearly 
corresponding to the star-forming concentrations traced by the H$\alpha$ 
emission. No evidence is seen for any violent nuclear activity.
 
Some weak extensions in all four quadrants are seen, best visible in
the 8.46 and 4.86\,GHz maps (Fig.~\ref{tp3} and Fig.~\ref{tp6_20} left).
The extensions become obvious after some median filtering is applied
(see Fig.~5 in T\"ullmann et al.~\cite{tul00}).
 
A very extended radio halo, which can be traced up to more than 1\arcmin{} from the galactic disk is seen in all maps. Assuming the distance to the galaxy of 26.7\,Mpc (Dahlem et al.~\cite{dahl95}) this corresponds to about
8\,kpc. We want to stress here, that this is the extension of the halo
as visible given by the noise-level of our observations and not the scale
height of the radio emission, which is, of course, smaller (see Sect.~\ref{sec:scale}).

Flux integrations in elliptical rings corresponding to the shape of the galaxy
give total flux densities of NGC\,5775 of $58 \pm 3$ mJy at 8.46\,GHz,
$96 \pm 5$ mJy at 4.86\,GHz and $257 \pm 13$ mJy at 1.49\,GHz.
These values fit well a single power-law with a spectral index of
$\alpha=0.85\pm0.02$ (where $S_\nu\propto\nu^{-\alpha}$), which is in agreement with earlier works (Duric et al.~\cite{dur98}). 
As the spectrum looks a bit flattened at the highest frequency we
estimated a mean nonthermal spectral index  of $\alpha_{nt}=0.92$ and,
correspondingly, a mean thermal fraction of  $\sim$9\% at 1.49\,GHz,
rising to $\sim$15\% at 8.46\,GHz.
 
\subsection{High-resolution polarization data}
 
\begin{figure}[htbp]
\resizebox{\hsize}{!}{\includegraphics[bb=40 200 560 680]{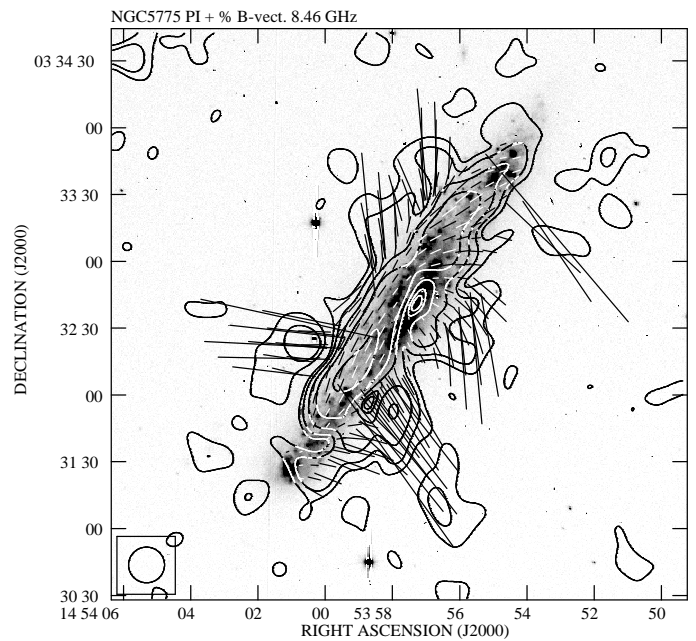}}
\caption[]{
Polarized intensity contour map at 8.46\,GHz with apparent 
polarization B-vectors, superimposed on the H$\alpha$ image
(from T\"ullmann et al.~\cite{tul00}).
Contours are plotted at levels (3, 5, 8, 12, and 20)
$\times$ 9\,$\mu$Jy/b.a. (r.m.s.). Vectors are proportional to
the degree of polarization (10$\arcsec$ corresponds to 10\%) and show
the apparent B-orientation.
Only vectors above 3$\sigma$ r.m.s. level (in both total
and polarized emission) are shown.
The resolution is 16\arcsec{} HPBW.
\label{pi3}
}
\end{figure}

\begin{figure}[htbp]
\resizebox{\hsize}{!}{\includegraphics[bb=40 200 560 680]{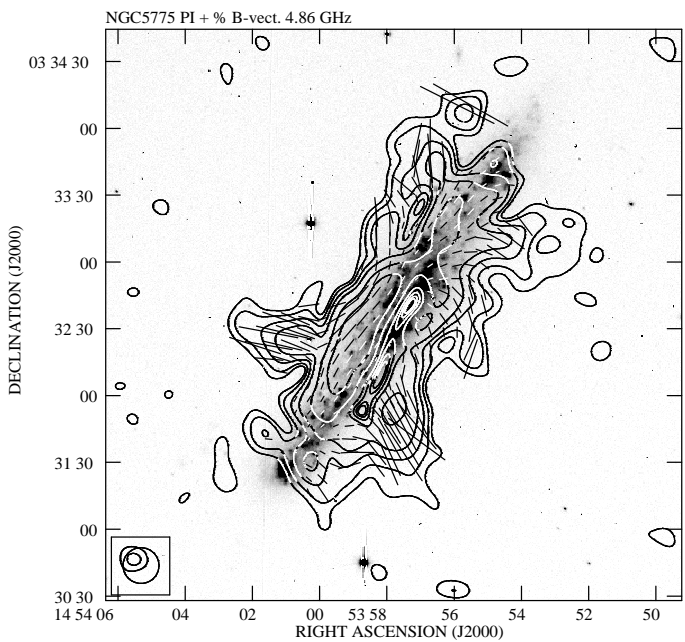}}
\caption[]{
Polarized intensity contour map at 4.86\,GHz with intrinsic (Faraday-corrected)
magnetic field orientation, superimposed on the H$\alpha$ image
(from T\"ullmann et al.~\cite{tul00}).
Contours are plotted at levels (3, 5, 8, 12, 20 and 30)
$\times$ 7\,$\mu$Jy/b.a. (r.m.s.). Vectors are proportional to
the degree of polarization (10$\arcsec$ corresponds to 10\%).
Vector plotting was suppressed below 3$\sigma$ r.m.s. level (in both total
and polarized emission).
The resolution is 16\arcsec{} HPBW.
\label{pi6}
}
\end{figure}

Figure~\ref{pi3} shows the polarized intensity (contours) with vectors of 
apparent magnetic field orientation at 8.46\,GHz and proportional to the
degree of linear polarization (and overlaid again on the H$\alpha$ map).
The contours of linearly polarized intensity at 4.86\,GHz are shown
in Fig.\ref{pi6},
together with polarization B-vectors corrected for Faraday rotation
(see also Sect.~\ref{sec:bfieldstructure}).
The degree of polarization does not exceed 10\% close to the 
disk plane and reaches 40\% in the outskirts of the galaxy.
 
The extensions in the four quadrants as described in Sect.~\ref{sec:tp}
are clearly visible in the polarized intensity maps, especially at
4.86\,GHz (Fig.\ref{pi6}) as this map has the highest signal-to-noise ratio.
The polarized intensity seems to form an X-shaped structure.
 
The apparent magnetic field orientation at 8.46\,GHz (Fig.~\ref{pi3}) close
to the galactic disk is plane parallel on the north-eastern side of the disk.
Vectors are turning rapidly (within a single beam-size area) when crossing the
disk and become almost perpendicular on the other side of the disk plane.
Such a configuration gives rise to a narrow beam-depolarized canal along
the south-western edge of the galactic disk
(from R.A.:\,$\mathrm{14^h53^m59^s}$; Dec.:\,$+03\degr32\arcmin00\arcsec$
to R.A.:\,$\mathrm{14^h53^m57^s}$;
Dec.:\,$+03\degr32\arcmin50\arcsec$). The depolarized canal is even better seen
at 4.86\,GHz (Fig.~\ref{pi6}) and not seen at 1.49\,GHz due to the overall
weak polarized signal at this frequency.
 
The extensions in polarized intensity and total power are accompanied with
vectors of apparent B-configuration that are also oriented at an angle of
$\sim$ 45\degr{} with respect to the galactic plane, forming the $X$-shaped
structure. The coincidence of polarized intensity 
with total power extensions is best visible in the higher frequency
maps (Fig.~\ref{tp3} and Fig.~\ref{tp6_20} left panel) but is still
indicated at the noise limit in the 1.49\,GHz map (Fig.~\ref{tp6_20} right).
 
The apparent magnetic field configuration is very similar at 8.46\,GHz and
4.86\,GHz (Fig.~\ref{tp3} and Fig.~\ref{tp6_20} left). This already allows us
to conclude that Faraday effects are not very strong in NGC\,5775 at these
short wavelengths. At 1.49\,GHz the polarized emission is detected in isolated
patches only and if compared to the higher frequency maps the observed
polarization orientation at this frequency is clearly Faraday-rotated
close to the disk plane. However, in the outskirts of the galaxy, the
apparent magnetic vectors even at 1.49\,GHz (where detected) agree well
with those at higher frequencies, suggesting only small Faraday effects there.
 
%=======================================
%===========================================
 
%=======================================
%===========================================
 
\section{Discussion}
\subsection{Total power emission and vertical scale heights}
\label{sec:scale}
 
\begin{figure}[hbtp]
\begin{center}
\resizebox{\hsize}{!}{\includegraphics{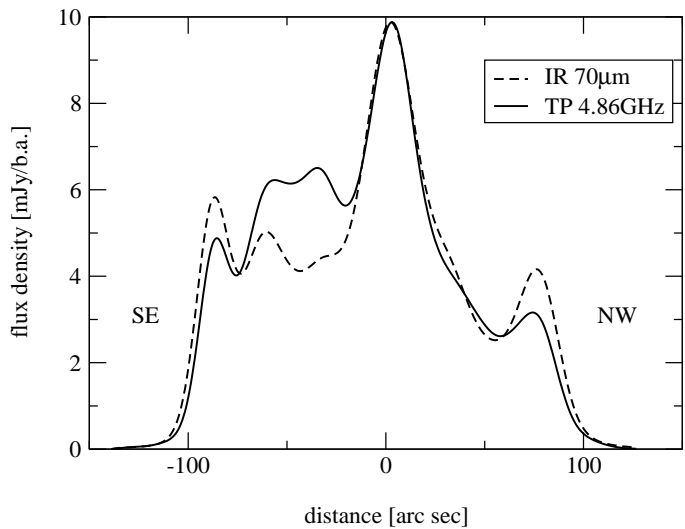}}
\end{center}
\caption[]{
Total intensity distribution at 4.86 GHz (16\arcsec{} HPBW) along the major
axis of NGC\,5775. The x-axis gives the distance from
the nucleus of the galaxy.
\label{majaxis}
}
\end{figure}

The intensity distribution of the total power along the major axis of the 
galaxy  at 4.86 GHz (Fig.~\ref{tp6_20}, left) is shown
in Fig.~\ref{majaxis}. 
 
%=======================================
%===========================================
 
The central maximum is slightly shifted to the northern part of the
galaxy with a steep decrease further north, but a slower decrease with
a plateau along the southern part of the major axis. The outermost local
maxima on both sides of the major axis are roughly at the same distance from
the nucleus ($r = 80 \arcsec$ or 10.7\,kpc) and probably indicate 
the positions of the outer spiral arms. The observed vertical extension
of the total power emission is due to the real disk thickness and the inclined
distribution of the radial disk emission, smoothed by the telescope beam.
In order to estimate the latter effect, we projected the radial total power
distribution (Fig.~\ref{majaxis}) to the inclination of the galaxy of
86\degr{} and convolved this distribution with a Gaussian function
with 16\arcsec{} HPBW to simulate the effect of the telescope beam.
The HPBW of the resulting distribution is 21.3\arcsec{} which is called
the \emph{effective} beam size. Only the parts of the emission profiles
perpendicular to the disk that exceed the effective beam size can be
attributed to extra-planar emission.
 
We determined the vertical scaleheights from emission profiles
that were obtained by strip integration perpendicular to the major
axis along 3 strips with a width of 60\arcsec{} (7.8\,kpc) each,
centered on the nucleus. These profiles were fitted with a model
distribution (for details see Dumke et al.~\cite{dum95}) consisting
of an intrinsic two-component exponential profile convolved with
the effective beam size. The fits were made  separately for the
emission 'above' (north = n) and 'below' (south = s of) the disk
midplane and at all three wavelengths. 

The values for both the thin and thick disks in n3 are significantly
larger than in the other strips, and also those in s3 deviate strongly
from the values in strips 1 and 2. According to T\"ullmann et
al.~(\cite{tul06}) there is most likely a background galaxy cluster
along the line of sight through strip n3, and possibly also along 
strip s3. This may influence the measured values at these
positions, so that we cannot regard them as being scaleheights of NGC\,5775.

The results for strips 1 and 2 are summarized
in Tab.~\ref{scale} where the strip number 1 refers to a radius
$-$90\arcsec{} to $-$30\arcsec{} and strip number 2 to $-$30\arcsec{} to
30\arcsec{} along the major axis (from south-east to north-west).

%====================================
\begin{table}[htbp]
\caption{Vertical scale heights for the thin and thick disk}
\label{scale}
\begin{center}
\begin{tabular}[]{crcrcrc}
\hline
&\multicolumn{2}{c}{8.46\,GHz}&\multicolumn{2}{c}{4.86\,GHz}
&\multicolumn{2}{c}{1.49\,GHz}\\
strip & $\rm h_{thin}$ & $\rm h_{thick}$ & $\rm h_{thin}$ &
$\rm h_{thick}$ & $\rm h_{thin}$ &$\rm h_{thick}$\\
&[pc]&[kpc]&[pc]&[kpc]&[pc]&[kpc]\\
\hline
n1 & 250 & 3.4 & 240 & 1.9 & 238 & 2.0\\
n2 & 301 & 2.4 & 202 & 1.7 & 228 & 1.9\\
s1 & 243 & 2.7 & 237 & 2.2 & 268 & 1.7\\
s2 & 320 & 1.8 & 295 & 2.0 & 286 & 1.5\\
\hline
mean & 279 & 2.6 & 244 & 2.0 & 255 & 1.8\\
& $\pm 33$ & $\pm 0.6$ & $\pm 33$ & $\pm 0.2$ & $\pm 24$ & $\pm 0.2$ \\
\hline
\end{tabular}
\end{center}
\end{table}
%====================================

The averages of the scaleheights in Tab.~\ref{scale} are given as
mean with errors reflecting the variations of the
scaleheights between the single strips. The 4.86\,GHz values for the
thin ($240 \pm 30$\,pc) as well for the thick disk ($2.0 \pm 0.2$\,kpc)
agrees within their errors with the scaleheights observed in four other
edge-on galaxies at this wavelength (Dumke \& Krause~\cite{dum98},
Heesen et al.~\cite{hee09a}). This sample of four galaxies includes
NGC\,253 with the brightest known halo and a very high star formation
rate (SFR) as well as one with the weakest halo, NGC\,4565, with
a small SFR. The star formation in NGC\,5775 is comparably strong
as that of NGC\,253. The observed scaleheight in NGC\,5775 supports
our previous result that a strong star formation does not increase
the global scaleheight of the radio emission (Krause~\cite{kra09}).
 
As shown in Tab.~\ref{scale} the thin-disk vertical scaleheights as 
well as the scaleheights for the thick disk agree with each other within 
their errors. Synchrotron losses decrease the vertical scale heights with 
increasing frequencies. On the other hand, the thermal fraction increases 
with frequency. The observed similarity of scaleheights at all three 
frequencies may be due to a superposition of both effects which also implies
that the thermal emission is not confined to the thin disk.

The analysis of the \emph{radial} dependence of the scaleheight at one
frequency of NGC\,253 showed indeed (together with spectral aging effects)
a vertical cosmic ray (CR) transport from the disk into the halo
(Heesen et al.~\cite{hee09a}). There, the CR lifetime is dominated by
synchrotron losses which depend on the magnetic field strength and is highest
in the central region. This leads to smallest scaleheights in the central
part and increasing values for larger radii visible as a dumbbell shape
of the halo. This effect is also visible in NGC\,5775: strip1 has larger
values for the thick disk than strip2 (Tab.~\ref{scale}). However,
the smaller angular extent of NGC\,5775 (which could not be compensated
by higher angular resolution of the observations) compared
to NGC\,253 and the contamination by the background galaxy cluster in
strip3 made a further analysis impossible.
 
\subsection{Spectral index distribution}
\label{sec:alpha}

\begin{figure}[hbtp]
\resizebox{\hsize}{!}{\includegraphics[bb=40 170 560 700]{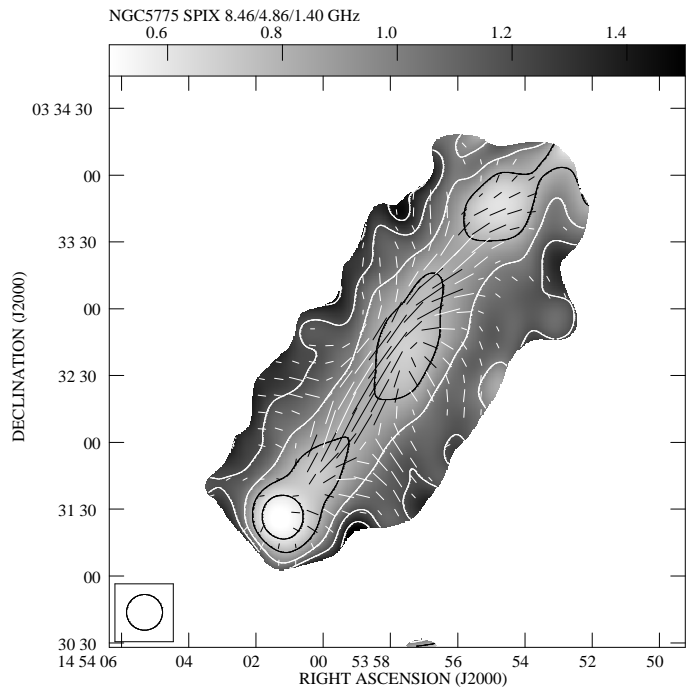}}
\caption[]{
Spectral index map calculated using all the data at  8.46\,GHz,
4.86\,GHz and  1.49\,GHz. Contours are of the same quantity, plotted at
$\alpha = 0.6$, 0.8 (black) and 1.0, 1.2, 1.4 (white) levels
($S_\nu=S\nu^{-\alpha}$). The resolution is 16\arcsec{} HPBW.
Magnetic vectors proportional to the polarized intensity at 4.86\,GHz
are overplotted (10\arcsec{} corresponds to 100\,$\mu$Jy/b.a.). 
\label{spix}
}
\end{figure}

The spectral index distribution was calculated using all three total
intensity maps as a fit of a power-law spectrum to each point
and is shown in Fig.~\ref{spix} (together with polarization
B-vectors at 4.86\,GHz). The spectrum is flattest along the disk
and especially on H$\alpha$ concentrations ($\sim$0.5) and steepens
smoothly (up to $\sim$1.3) towards the outer boundary of the halo. 
The steepening seems to be slightly stronger towards the
north-eastern boundary ('above' the disk) than towards the
south-western boundary ('below' the disk) where even a slight
flattening is indicated just west of the nucleus which coincide with
extended X-ray emission as shown in T\"ullmann et al.~(\cite{tul06}). 
Some steepening of the spectrum can be noticed along with all
four polarized intensity extensions, which may reflect slightly faster
synchrotron losses in the locally enhanced magnetic field there. 

Our spectral index map made of three frequency observations cannot
confirm the vertical structures (chimneys) of flattened spectral
indices as presented by Duric et al. (\cite{dur98}) who determined
the spectral index map from the VLA observations at 4.86\,GHz and
1.49\,GHz alone. This can hardly be due to the slightly higher
angular resolution of their map ($15\arcsec \times 13\arcsec$)
compared to our $16\arcsec$~HPBW map but, as we use a larger dataset
(all available in the VLA archive), hence it may be artifacts in their maps.
 
%=======================================
%===========================================
 
\begin{figure}[hbtp]
\begin{center}
\resizebox{\hsize}{!}{\includegraphics{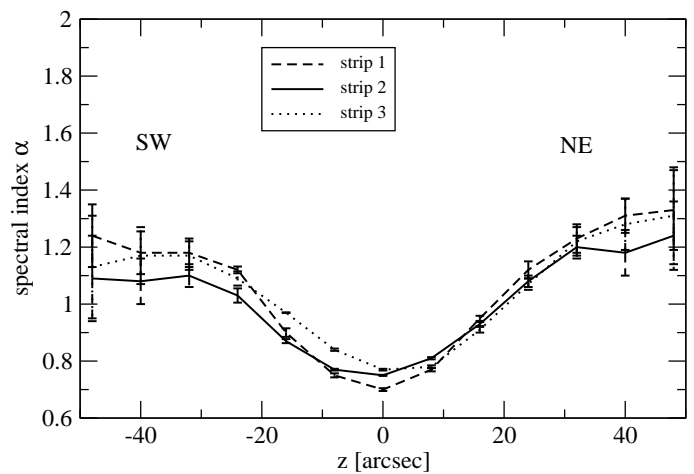}}
\end{center}
\caption[]{
Spectral index distribution within three strips perpendicular
to the major axis: strip~1 (dashed) refers to a radius $-$90\arcsec{}
to $-$30\arcsec, strip~2 (full) from $-$30\arcsec{} to 30\arcsec, and
strip~3 (dotted) from 30\arcsec{} to 90\arcsec{} along the major axis
(from south-east to north-west). 
\label{alpha}
}
\end{figure}

In order to quantify the spectral index variations we determined
the spectral index between all 3 frequencies in the same 3 strips
described in Sect.~\ref{sec:scale}. The values are presented in
Fig.~\ref{alpha}. The spectra increase in all strips from the
midplane with values of about $\alpha = 0.75$ up to $\alpha = 1.1$
at 32\,arcsec{} distance on both sides away from the midplane.
Such a steepening is expected by a decrease of the thermal fraction
from the midplane to higher z-values. Additionally, synchrotron
losses of the nonthermal radiation are expected to increase with
increasing distance from the midplane. As the highest reliable
values for the spectral indices are about $\alpha = 1.1$ and the
value for the mean nonthermal spectral index  $\alpha_{nt}=0.92$ (as
estimated in Sect.~\ref{sec:tp}). Assuming a thermal content in the disk
of about 11\% at 1.49\,GHz (equivalent to about 35\% at 8.46\,GHz),
we can explain the spectrum flattening from the mean $\alpha_{nt} = 0.92$
down to $\alpha = 0.75$.
The further steepening -- up to $\alpha = 1.1$ -- we can attribute to
synchrotron losses at the galactic peripheries.
 
%=======================================
%===========================================
 
The spectral index distribution in strip 3 (dotted) is slightly
shifted to positive z-values with respect to the distributions in
the two other strips. This shift seems to be significant as the
errors are very small in the central parts. It may be due to the
background galaxy cluster in strip3 as discussed in Sect.~\ref{sec:scale}. 
 
Even in the strip integration we see an indication of a spectral
index flattening in the central strip (full line) for
z~$\le -30 \arcsec$ ('below' the disk) where the X-ray emission is strongest
(T\"ullmann et al.~\cite{tul06}).

\subsection{Vertical distribution of the polarized intensity}
 
\begin{figure}[hbtp]
\resizebox{\hsize}{!}{\includegraphics{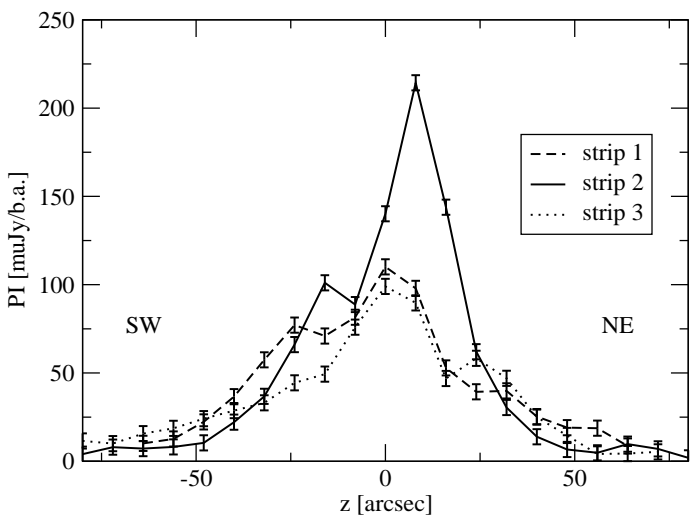}}
\caption[]{
Polarized intensity at 4.86\,GHz averaged along three strips of
60\arcsec{} width each, perpendicular to the major axis: strip~1
(dashed) refers to a radius $-$90\arcsec{} to $-$30\arcsec, strip~2
(full) from $-$30\arcsec{} to 30\arcsec, and strip~3 (dotted)
from 30\arcsec{} to 90\arcsec{} along the major axis (from south-east
to north-west). 
\label{stripPI}
}
\end{figure}

As described in Sect.~\ref{sec:tp} the region with the magnetic
field plane parallel to the galactic disk is slightly shifted
to the north-east ('upper') part of the disk (best visible in
Fig.~\ref{tp3}). We wanted to test whether this asymmetry in
the distribution of magnetic field orientation is also accompanied
by an asymmetry in the distribution of the polarized intensity.
Therefore we integrated also the polarized intensity at
4.86\,GHz in the same three strips parallel to the minor axis
as described in Sect.~\ref{sec:scale}. The result is shown in
Fig.~\ref{stripPI}.
 
%=======================================
%===========================================
 
The maximum of the polarized intensity of the central strip (full line)
is clearly shifted to positive z-values (north-east), the maxima
in the two other strips are slightly shifted towards the same direction.
All maxima agree in position with the regions of the plane-parallel
magnetic field orientation. Further, all distributions of polarized
intensity are asymmetric in shape: they show a rather strong
decrease with increasing $|z|$, but have a shoulder towards negative
z-values where the magnetic field orientation is mainly vertical
to the galactic plane. 
 
From X-ray observations (e.g. T\"ullmann et al. \cite{tul06}) we
can conclude that the north-eastern side of NGC\,5775 is the near
side, hence we see the disk slightly from 'below'. This can explain
why we see the disk-parallel magnetic field which is supposed
to be strongest, somewhat shifted to the north-east with respect
to the galactic midplane (see discussion in Sect.~\ref{sec:bfieldstructure}).
 
\subsection{Faraday rotation and depolarization}
\label{sec:RM}
 
\begin{figure}[htbp]
\resizebox{\hsize}{!}{\includegraphics[bb=40 170 560 715]{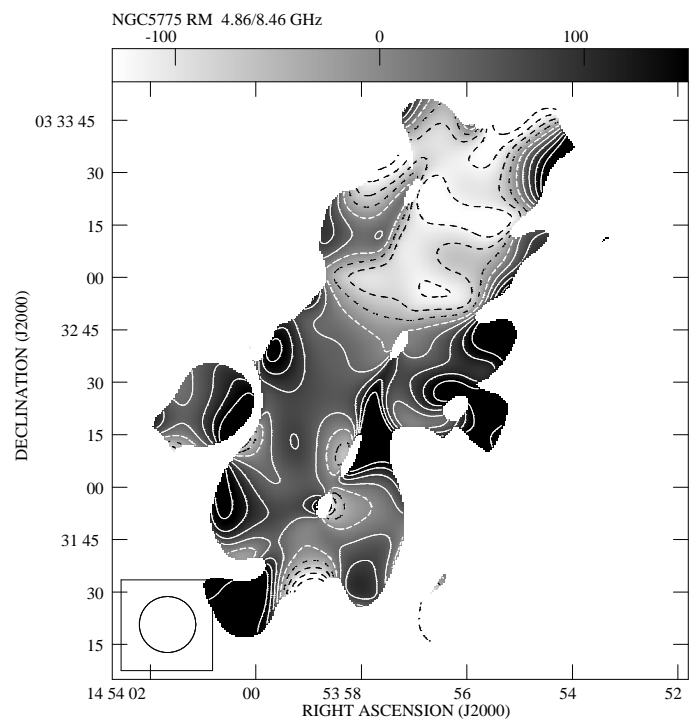}}
\caption[]{
The Faraday rotation measure distribution determined between 8.46\,GHz
and 4.86\,GHz maps. Contours are plotted in RM intervals of 40\,rad/m$^2$
-- the dashed white contour corresponds to the RM value of 0\,rad/m$^2$.
The resolution is 16\arcsec{} HPBW.
\label{rm}
}
\end{figure}

The Faraday rotation measure distribution, calculated between 8.46\,GHz
and 4.86\,GHz, is presented in Fig.~\ref{rm} in gray-scales with
contours -- white and solid for positive values, black and dashed for 
negative ones. The distribution was determined from data
truncated at the 3 $\sigma$ (r.m.s.) levels in the polarized intensity maps
resulting in a maximal uncertainty at the limits of detected polarized 
signal of about 100\,rad/m$^2$.
 
The overall rotation measure (RM) has values between $\pm$140\,rad/m$^2$ 
and exceeds this range only in small isolated areas where uncertainties
are substantial. Its distribution is well balanced among positive and
negative values. This suggests a negligible foreground Faraday
rotation at the position of NGC\,5775. This is in agreement with
measurements of, e.g., Taylor et al. (\cite{tay09}).

There is a clear trend along the major axis: RM is negative
in the north-western and positive in the south-eastern half of the galaxy.
Two jumps in RM distribution (close to position of
R.A.:\,$\mathrm{14^h53^m58^s}$; Dec.:\,$+03\degr32\arcmin00\arcsec$) are 
caused by large uncertainty of RM determination in locations with weak 
polarized intensity (compare with polarized intensity maps, particularly
at 4.86\,GHz in Fig.\,\ref{pi6}).
 
%=======================================
%===========================================
 
\begin{figure}[htbp]
\resizebox{\hsize}{!}{\includegraphics[bb=40 160 560 710]{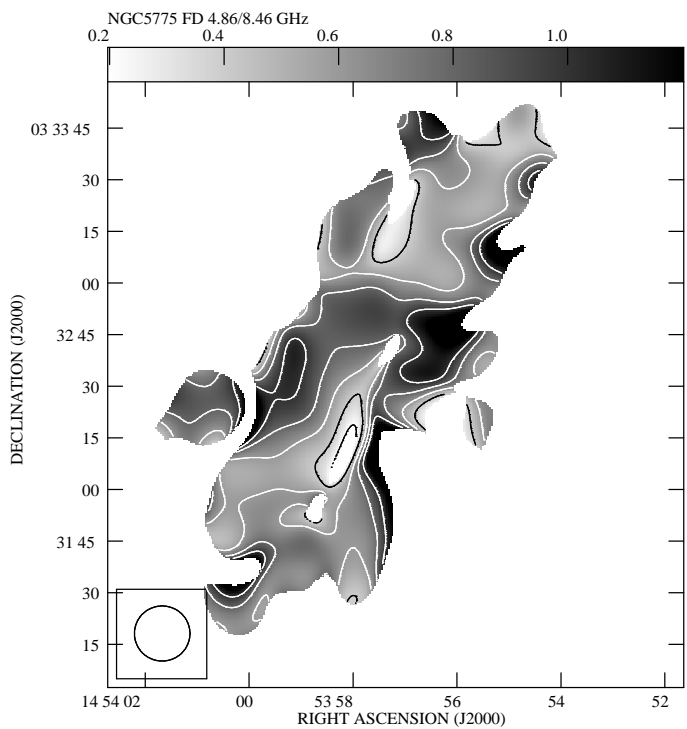}}
\caption[]{
Depolarization map calculated as the ratio of polarization
degree at 4.86\,GHz to that at 8.46\,GHz. Contours are of the same
quantity, plotted at 1.0, 0.8, 0.6 (white) and 0.4 and 0.2 (black)
levels. The resolution is 16\arcsec{} HPBW. The data were clipped 
at the 3$\sigma$ level in total and polarized intensity prior to the
determination of this this quantity.
\label{fd}
}
\end{figure}

The Faraday depolarization was determined 
as the ratio of the polarization degree at 4.86\,GHz to that at 8.46\,GHz
and is presented in Fig.~\ref{fd}. The data were clipped 
at the 3$\sigma$ level in total and polarized intensity prior to the determination
of this this quantity. We decided to use the data at the two higher frequencies 
since the 1.49\,GHz polarized emission exceeds the cutting level only in
very limited area.
The distribution is smooth, with mean values of 60\%--80\%.
Some axial symmetry is visible with higher values (less depolarization)
found close to the rotation axis in agreement with the Faraday
rotation distribution, as we expect stronger depolarization where
the Faraday effects (rotation and dispersion) are larger. 
The depolarization is smallest (in absolute values)
close to the minor axis growing in both directions along the major axis.
Again, the area with low polarized signal (at about
R.A.:\,$\mathrm{14^h53^m58^s}$; Dec.:\,$+03\degr32\arcmin00\arcsec$)
makes the depolarization very high there (low values),
but -- as in case of RM jumps -- with large uncertainties.
 
As mentioned before, we inferred from the position of the dust lane 
and the X-ray distribution
(T\"ullmann et al.~\cite{tul00} and \cite{tul06}) that the
north-eastern side is the near side and the south-western is the far side
of NGC\,5775. Hence, we expect that the emission along the south-western
side of the major axis is more affected by Faraday rotation and
depolarization effects than the north-eastern region along the major axis,
which may explain the observed asymmetry in RM and depolarization
perpendicular to the major axis.
 
%=======================================
%===========================================
 
\subsection{Magnetic field strength}
\label{sec:bfieldstrength}
 
The magnetic field strength was calculated from the 4.86\,GHz data,
assuming energy equipartition between cosmic-rays and magnetic fields (see
Beck \& Krause~\cite{bfeld}). We assumed a ratio of proton-to-electron
energy density of 100.
For a non-thermal spectral index in the extensions we assumed the
value of $\alpha$=1.1, which corresponds to the
value in the spectral index map (Fig.~\ref{spix}) there, and a mean 
spectral index of $\alpha_{nt}$=0.92 (as derived in Sect.~\ref{sec:tp})
for the determination close to the disk plane.
Assuming an axisymmetric magnetic field in the disk, 
close to the galactic disk we obtained values for the total and ordered
magnetic field of $B_t=11\pm3$\,$\mu$G, $B_{reg}=4\pm1$\,$\mu$G, respectively. 
In all four $X$-shaped extensions, we determined $B_t=6\pm2$\,$\mu$G,
$B_{reg}=4\pm1$\,$\mu$G. The total path length through the emitting 
volume was adopted as 20\,kpc through the galactic midplane (see Fig.~\ref{majaxis}) and as 8\,kpc through the extensions. The magnetic field uncertainty takes into account uncertainties by a factor of two of the proton-to-electron ratio and the disk thickness.
 
With these values, the magnetic field strength in NGC\,5775 is
comparable to the field strength in NGC\,253 (Heesen et al. \cite{hee09a}), which also hosts a nuclear starburst and strong star formation in the disk.
 
\subsection{Magnetic field structure}
\label{sec:bfieldstructure}
 
The observed `apparent' magnetic vectors need to be corrected for Faraday
rotation in order to derive the intrinsic magnetic field orientation.
The Faraday rotation in NGC\,5775 is $\le\pm30\degr$ ($|RM| \le 140$\,rad/m$^2$)
at 4.86\,GHz. The corrected vectors are shown in Fig.~\ref{pi6}.
The orientations of the vectors is very similar to the observed B-vectors
at 8.46\,GHz as the Faraday rotation is almost negligible ($\le 10 \degr$
for $|RM| \le 140$\,rad/m$^2$) at this high frequency. 
 
Close to the disk the magnetic orientation is plane-parallel.
Magnetic vectors above 1\,kpc from the galactic disk, however, are
dominated by vertical components forming an X-shaped structure
(Fig.~\ref{pi6}) as has been observed in several other edge-on galaxies
(see e.g. Krause~\cite{kra09}).
This pattern of the magnetic field vectors (presenting the direction
of the magnetic field component projected on the sky-plane) together
with the smooth and symmetric distribution of the RM distribution
(Fig.~\ref{rm}) as a measure of the magnetic field strength 
along the line of sight shows that NGC\,5775 hosts an ordered large-scale
magnetic field configuration. 

\begin{figure}[htbp]
\resizebox{\hsize}{!}{\includegraphics{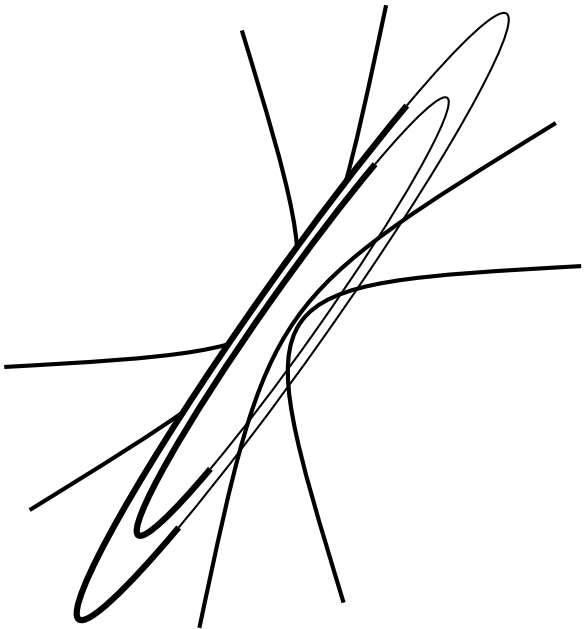}}
\caption[]{
Sketch of the large-scale magnetic field configuration in the disk and in the halo.
\label{sketch}
}
\end{figure}

Figure~\ref{sketch} shows a sketch of the two predominant components of the
large-scale magnetic field in NGC\,5775, the plane-parallel ASS disk field and the
X-shaped halo field. The disk component seems to be highly asymmetric in its
regularity. The maximum of polarized intensity is on the near side, shifted
to the north-east and we see mostly emission from this field component.
On the far side the magnetic field regularity is smaller and more of
the synchrotron emission from the halo component can be seen.
This situation is drawn in Fig.~\ref{sketch}, where the line thickness
illustrates the disk field regularity.

Such an asymmetry in magnetic field regularity is commonly seen in radio
polarimetric observations of interacting galaxies (e.g. NGC\,4254 Chy\.zy et
al.~\cite{chy02}). Indeed, NGC\,5775 is known to interact with NGC\,5774,
a faint dwarf galaxy projected to the north-west of NGC\,5775
(Irwin~\cite{irwin94}, Gosh et al.~\cite{gosh09}). There are two \ion{H}{i}
bridges with gas transfer from NGC\,5774 to NGC\,5775. Optical and radio
emission was also detected along these bridges and Irwin~\cite{irwin94}
suggested that this system may be in an \emph{early} stage of a merger.
This was supported by a recent analysis of X-ray data by Gosh et
al.~\cite{gosh09} and may explain why the magnetic field regularity
is still high along one side of NGC\,5775.
 
\subsection{Dynamo action and galactic wind}
\label{sec:wind}
 
The optical spiral pattern in galaxies is usually trailing, and the magnetic
field orientation is primarily parallel to the optical spiral arms as observed
in all face-on spiral galaxies studied so far. If the direction of the disk
rotation is known, one can use the RM-distribution to determine wether
the magnetic field direction is inwards or outwards (Krause et
al.~\cite{kra89}). The kinematics of NGC\,5775 is well studied
(e.g. Irwin~\cite{irwin94}) with the north-western side 
approaching and the south-eastern side receding. With the RM being
predominantly negative on the approaching side we can conclude that
the direction of the large-scale magnetic field in NGC\,5775 is \emph{outwards}. 
Together with NGC\,4254 (Chy{\.z}y~\cite{chy08}) and NGC\,891
(Krause~\cite{kra09}) there are now three galaxies found with outwards
directed ASS fields. The other four galaxies for which the direction of
the ASS could be determined up to now (M\,31, IC\,342, NGC\,253, and NGC\,6949)
were all inwards directed, which could not be explained up to now
(Krause \& Beck~\cite{krause98}). The three counterexamples NGC\,891, NGC\,4254,
and NGC\,5775 balance the statistics as expected by theory which does not
predict a preferred field direction.

A large-scale (unidirectional) magnetic field in a disk of the galaxy
is generally thought to be dynamo-generated, e.g. by the mean
field $\alpha\Omega$-dynamo.
The magnetic field configuration that is most easily excited by this
dynamo is the axisymmetric field (ASS) (e.g. Ruzmaikin et al.~\cite{Ruz88}).
This field configuration consists of a strong unidirectional toroidal
component with magnetic field lines running azimuthally
within the disk and a weaker poloidal field with field lines looping
perpendicularly to the azimuthal direction.
The poloidal field can either be of quadrupole (separate loops above and
below the plane -- \emph{even}-type) or dipolar (\emph{odd}-type) symmetry
(loops across the plane). In the even ASS-mode the magnetic field is
symmetric with respect to the galactic mid-plane (z=0) which is not the
case for the odd mode (see Fig.~9 in Krause et al. \cite{kra89}).
Hence, in the even mode, the direction of the ASS disk field in
the galactic disk is equal above and below the disk plane whereas
in the odd mode it has opposite directions on both sides of the
mid-plane.

For our case, the important consequence is the rotation
measure signature for both modes, as seen in nearly edge-on disks.
The even mode shows the antisymmetry with respect to the minor axis
in the RM distribution and symmetry according to the major one. 
The odd mode shows antisymmetry with respect to both axes.
As seen in Fig.~\ref{rm}
and as discussed in Sect.~\ref{sec:RM} the RM distribution has a 
clear asymmetry \emph{along} the galactic mid-plane and no change of its sign
across the plane. This and the fact that the field lines are oriented
along the mid-plane agrees with an \emph{even} axisymmetric disk
field in the disk of NGC\,5775 with magnetic field lines directed at the front
side from south-east to north-west. 
 
The accompanying quadrupolar poloidal field of this even ASS field
configuration in the disk alone cannot explain the observed X-shaped
field structure in NGC\,5775, as it is -- according to the mean-field dynamo
theory -- by a factor of about 10 weaker than the azimuthal disk field.
However, model calculations of the mean-field $\alpha\Omega$-dynamo
for a disk surrounded by a spherical halo including a {\em galactic wind}
(Brandenburg et al.~\cite{bran93} and Moss et al.~\cite{moss10})
simulated similar field configurations as the observed ones.
Meanwhile, new MHD simulations of disk galaxies including
a galactic wind are in progress and may explain the X-shaped field
(Gressel et al. \cite{gress08}; Hanasz et al. \cite{MH09a,MH09b}).
The first global galactic-scale MHD simulations of a CR-driven dynamo
give very promising results showing directly that magnetic flux
is transported from the disk into the halo (Hanasz et al. \cite{MH09c}). 

Even for face-on galaxies with inclinations less than 60\degr,
Braun et al.~(\cite{brau10}) found evidence for a large-scale halo field
from polarization asymmetries along the major axis which are consistent
with a quadrupolar halo field and not with a dipolar halo field.
As a quadrupolar halo field shows -- at least in its inner part -- a similar
structure as an X-shaped magnetic field (which is different from a dipolar
halo field) their results for face-on galaxies fit to our findings of
X-shaped halo fields in edge-on galaxies.

The kinematics of ionized gas as well as the X-ray properties (Lehnert \&
Heckman \cite{LH96}, T\"ullmann et al. \cite{tul06}) suggest that
NGC\,5775 is even a ``super-wind'' galaxy. Such strong galactic winds
from violent star formation areas should give rise to an almost
spherical halo.

The large scale distribution of the various gas phases in the halo of
NGC\,5775 and its large scale magnetic field distribution is reminiscent of
other galaxies with evidence for star-burst driven winds such as NGC\,253
(e.g. Heesen et al. 2009a,b) or NGC\,4666 (e.g. Dahlem et al. 1997). The X-ray
halo of NGC\,5775 is well studied and its properties (T\"ullmann et al. 2006, 
Li et al. 2008) corroborate the scenario for a ``super-wind'' from the inner
disk (Strickland \& Stevens 2000) being present in this object.
The examples of this class of objects mentioned above also have in common
that the large scale magnetic field orientation in the halo has
a characteristic X-shape. The distribution of the synchrotron radiation
and its polarization characteristics result from the combination of the wind
affecting the cosmic ray propagation and the magnetic field structure.
The X-shaped structure of the wind results naturally from
the centrifugal barrier of the effective galactic potential as has been
shown by wind models for disk galaxies (e.g., Sushkov et al. 1998,
Dalla Vecchia \& Schaye 2008).
Unfortunately, all wind models so far neglect the influence of the magnetic field, which can only be included by full MHD simulations. A galactic wind may generally play an important role in the amplification of the large-scale magnetic field by the mean-field dynamo, as it may solve the helicity problem of dynamo action (e.g. Sur et al.~\cite{sur07}).
 
\section{Summary and conclusions}
 
Deep polarized radio-continuum observations of the violently star forming 
galaxy NGC\,5775 at 8.46\,GHz together with archive data at lower frequencies
allow us to reveal its global magnetic field configuration.
We determined the total flux densities at all frequencies which
fit well a single power-law with a spectral index of $\alpha=0.85\pm0.02$.
 
The rotation measure distribution varies smoothly on both sides along
the major axis from positive values in the south-east to negative values
in the north-west. The intrinsic magnetic field orientation along
the galactic midplane is plane-parallel. From this we conclude that
NGC\,5775 has an \emph{even axisymmetric} (even ASS) large-scale magnetic
field configuration in the disk as generated by an $\alpha\Omega$-dynamo
which is accompanied by a quadrupolar poloidal field. Taking into account
informations about rotation of NGC\,5775 we can conclude that the
radial component of the plane-parallel magnetic field points \emph{outwards}.
 
The intrinsic magnetic field pattern further away from the galactic plane 
forms an X-shaped structure as known from observations at lower
frequency. Not only the magnetic field lines are X-shaped but also
the distribution of the polarized intensity is X-shaped. Four extensions
are even indicated in the total power maps being accompanied with a small
steepening of the spectral index.
 
The magnetic strengths were determined to be $B_t=12\pm3$\,$\mu$G for
the total and $B_{reg}=4\pm1$\,$\mu$G for the ordered field in the
midplane (the ASS field). The field strength in the X-shaped structure
reaches nearly the same values with $B_t=9\pm2$\,$\mu$G and
$B_{reg}=4\pm1$\,$\mu$G. The field strength of the observed X-shaped
halo magnetic field is by about one order of magnitude too high to
be explained by the action of a mean-field dynamo alone. It may be due,
however, to an interplay of the galactic wind in NGC\,5775 together with
dynamo action.
 
\begin{acknowledgements}
We thank the staff of the 100-m Effelsberg telescope for their assistance
with the radio observations. This work benefited from the exchange program
between the Ruhr-Universit\"at Bochum (RUB) and the Jagiellonian University
Krak\'ow. Our thanks go to numerous colleagues at RUB and Astronomical
Observatory of the Jagiellonian University for their valuable comments.
We would like to thank the referee for his effort in improving the article.
Support by grants from the Polish Government, grants no. PB3033/H03/2008/35
and 92/N-ASTROSIM/2008/0 as well as by Deutsche Forschungsgemeinschaft
in the framework of SFB591 and FOR1048 at RUB is greatfully acknowledged.
 
\end{acknowledgements}


\begin{thebibliography}{}
\bibitem[1977]{baa77}
        Baars J.W.M., Genzel R., Pauliny-Toth I.I.K., Witzel A.:1997 \aap 61, 99
\bibitem[2005]{bfeld}
        Beck R., Krause M.: 2005, AN 326, 414
\bibitem[1998]{birk98}
        Birk G.T., Lesch H., Neukirch T.: 1998, MNRAS 296 165
\bibitem[1993]{bran93}
        Brandenburg A., Donner K.J., Moss D., Shukurov A.M., Sokoloff D.D.,
        Tuominen I.: 1993, A\&A 271, 36
\bibitem [2010]{brau10} 
         Braun, R., Heald, G., Beck, R.: 2010, A\&A 514, 42
\bibitem[2002]{chy02}
        Chy{\.z}y K.T., Urbanik M., Soida M., Beck R.: 2002, Ap\&SS 281, 409
\bibitem[2008]{chy08}
        Chy{\.z}y K.T.: 2008, A\&A 482, 755
\bibitem[2000]{col00}
        Collins J.A., Rand R.J., Duric N., Walterbos R.A.M.: 2000, ApJ 536, 645
\bibitem[2001a]{CR01a}
        Collins J.A., Rand R.J.: 2001a, in {\it Galaxy Disks and
        Disk Galaxies}, ASP Conference Series, Vol. 230, eds. J.G.
        Funes,and E.M. Corsini. San Francisco: Astronomical Society
        of the Pacific p. 307
\bibitem[2001b]{CR01b}
        Collins J.A., Rand R.J.: 2001b, ApJ 551, 57
\bibitem[1995]{dahl95}
        Dahlem M., Lisenfeld U., Golla G.: 1995, ApJ 444, 119
\bibitem[2006]{da06} 
         Dahlem M., Lisenfeld U., Rossa, J.: 2006, A\&A 457, 121
\bibitem[1997]{Dahl97} Dahlem, M., Petr, M. G., Lehnert, M. D. et al.: 1997,
  A\&A 320, 731
\bibitem[2008]{Vecchia08} Dalla Vecchia, C., Schaye, J.: MNRAS 387,1431

\bibitem[1995]{dum95}
         Dumke M., Krause M., Wielebinski R., Klein U.: 1995, A\&A 302, 691
\bibitem[1998]{dum98}
         Dumke M., Krause M.: 1998, in {\it The Local Bubble and Beyond},
         eds. D. Breitschwerdt, M. Freyberg, J. Tr\"umper, Proc. IAU Coll.
         166, Lecture Notes in Physics 506, Springer-Verlag, Berlin, p.555
\bibitem[1998]{dur98}
        Duric N., Irwin J., Bloemen H.: 1998, A\&A 331, 428
\bibitem[1979]{EG88}
        Emerson D.T., Gr\"ave R.: 1988, A\&A 190, 353
\bibitem[1992]{els92}
        Elstner D., Meinel R., Beck R.: 1992, A\&ASS 94, 587
\bibitem[1994]{GH94}
        Golla G., Hummel E.: 1994, A\&A 284, 777
\bibitem[2009]{gosh09}
        Gosh K.K., Saripalli L., Gandhi P., Foellmi C., Guti\'errez C.M., L\'opez-Corredoira M.: 2009, ApJ 137, 3263
\bibitem[2008]{gress08}
        Gressel O., Ziegler U., R\"udiger G.: 2008, A\&A 486, L35
\bibitem[2003]{HL03}
        Hanasz M., Lesch H.: 2003, A\&A 412, 331
\bibitem[2009a]{MH09a}
        Hanasz M., Otmianowska-Mazur K., Kowal H., Lesch H.:2009a, A\&A 498, 335
\bibitem[2009b]{MH09b}
        Hanasz M., Otmianowska-Mazur K., Lesch H. et al.:2009b, in IAU Symp. Vol. 259, 479-484
\bibitem[2009c]{MH09c}
        Hanasz M., W\'olta\'nski D., Kowalik K.: 2009c, ApJ 706, L155
\bibitem[2009]{hee09a}
        Heesen V., Beck R., Krause M., Dettmar R.-J.: 2009a, A\&A 494, 563
\bibitem[2009]{hee09b}
        Heesen V., Krause M.,Beck R., Dettmar R.-J.: 2009b, A\&A 506, 1123
\bibitem[1994]{irwin94}
        Irwin J.A.: 1994, ApJ 429, 618
\bibitem[1998]{krause98}
        Krause F., Beck R.: 1998, A\&A 335, 786
\bibitem[1989]{kra89}
        Krause M., Hummel E., Beck R.: 1989, A\&A 217, 4
\bibitem[2009]{kra09}
        Krause M.: 2009, in Magnetic Fields in the Universe II, ed.
        A. Esquirel, Rev. Mex. Astron y Astrof. 36, 25
\bibitem[2001]{lee01}
        Lee S.-W., Irwin J.A., Dettmar R.-J., Cunningham C.T.,
        Golla G., Wang Q.D.: 2001, A\&A 377, 759
\bibitem[2002]{lee02}
        Lee S.-W., Seaquist E.R., Leon S., Garc\'ia-Burillo S.,
        Irwin J.A.: 2002, ApJ 573, L107
\bibitem[1996]{LH96}
        Lehnert M.D., Heckman T.M.: 1996, ApJ 472, 546
\bibitem[2008]{Li08} Li, J.T., Li, T., Wang, D., et al.: 2008, MNRAS 390, 59
\bibitem[2010]{moss10} Moss D., Sokoloff D., Beck R., Krause, M.: 2010, A\&A
	512, A61
\bibitem[2000]{rand00}
        Rand R.J.: 2000, ApJ 537, L13
\bibitem[2000]{RD00}
        Rossa J., Dettmar R.-J.: 2000, A\&A 359, 433
\bibitem[1988]{Ruz88}
        Ruzmaikin A.A., Shukurov A.M., Sokoloff D.D.: 1988,
        ``Magnetic Fields of Galaxies'', Kluwer, Dordrecht
\bibitem[2006]{}
        Shukurov, A., Sokoloff, D., Subramanian, K. et al.: 2006, A\&A
  448, L33
\bibitem[2000]{Strick00} Strickland, D. K., Stevens, I. R.: 2000, MNRAS 314, 511
\bibitem[2007]{sur07}
        Sur S., Shukurov A., Subramanian K.: 2007, MNRAS 377, 874
\bibitem[1996]{Such98} Suchkov, A. A., Berman, V. G., Heckman, T. M. et al.:
  1998, ApJ 463, 528 
\bibitem[2007]{tab07}
        Tabatabaei F.S., Beck R., Kr\"ugel E., Krause M.,
        Berkhuijsen E.M., Gordon K.D., Menten K.M.: 2007, A\&A, 475, 133
\bibitem[2009]{tay09}
	Taylor A.R., Stil J.M., Sunstrum C.: 2009, ApJ 702, 1230
\bibitem[2000]{tul00}
        T\"ullmann R., Dettmar R.-J., Soida M., Urbanik M., Rossa J.:
        2000, A\&A 364, L36
\bibitem[2006]{tul06}
        T\"ullmann R., Pietsch W., Rossa J., Breitschwerdt D., Dettmar R.-J.:
        2006, A\&A 448, 43
\end{thebibliography}
\end{document}